\documentclass[aps,prl,psfig,twocolumn,floats,showpacs]{revtex4}
\usepackage{psfig}

\begin{document} 

\title{Ultrafast dynamics of coherences 
in the quantum Hall system} 

\author{K.M. Dani$^{1,2}$, J. Tignon$^3$,  
M. Breit$^2$ and D.S. Chemla$^{1,2}$}

\affiliation{$^1$Department of Physics, 
University of California at Berkeley,}

\affiliation{
 $^2$Materials Sciences Division,
E.O. Lawrence Berkeley National Laboratory, Berkeley, California
94720,}

\affiliation{ $^3$Laboratoire Pierre Aigrain, 
Ecole Normale Sup\'erieure, F-75005 Paris, France.}

\author{E.G. Kavousanaki and  I.E. Perakis}

\affiliation{
Institute of Electronic Structure \& Laser, Foundation
for Research \& Technology-Hellas,
and Department of Physics, University of Crete, Heraklion, Greece.}

\date{\today}

\begin{abstract}

Using three-pulse four-wave-mixing optical spectroscopy, we study 
the ultrafast dynamics of the quantum Hall system. We observe 
striking differences as compared to an undoped system, where the 
2D electron gas is absent. In particular, we observe a large 
off-resonant signal with strong oscillations.
Using a microscopic theory, we show that these are 
due to many-particle coherences created by interactions between 
photoexcited carriers and collective excitations of the 2D electron 
gas. We extract quantitative information about the dephasing and interference 
of these 
coherences.

\end{abstract}


\maketitle 

%\section{Introduction}
%\label{sec:intro}

The quantum Hall effects \cite{QHE3,QHE1,QHE2} arise in a cold 
two-dimensional electron gas (2DEG) in a perpendicular magnetic 
field. They result from Coulomb correlations among the ground state 2DEG 
electrons that populate the highly degenerate Landau level states.
Previous transport and optics experiments have studied the 
properties of this incompressible electron fluid and its elementary 
excitations \cite{QHE1,aron}. Of interest here are the collective 
charge excitations called magnetoplasmons (MP). 
Unlike for zero magnetic field \cite{plasmon}, 
in the quantum Hall regime the  correlations 
and incompressible ground state result in a 
pronounced magnetoroton  minimum \cite{QHE2}.
Despite their large momenta,  magnetorotons dominate the 
inelastic light scattering \cite{QHE1,QHE2,aron}. 
However, such experiments cannot access MP dynamics. Moreover, they 
cannot access the early timescales required to observe {\em coherent} 
effects  in the quantum Hall system. Coherences 
play a central role in several quantum mechanical 
systems \cite{axt-04, bil,eit,lwi,inter-val}. 
Recent proposals for quantum computing point out the need to study 
the coherent dynamics 
of the quantum Hall system \cite{dassarma}.  
Ultrafast non-linear optics 
is just beginning to explore these phenomena 
\cite{from-02,kara,pera-05}.

Ultrafast four-wave-mixing (FWM) spectroscopy is well suited for 
studying coherent dynamics \cite{Ultra1}. It has demonstrated that 
Coulomb interactions in undoped semiconductor quantum wells are crucial 
to this dynamics and lead to exciton-exciton and carrier-phonon interactions, 
non-Markovian memory effects, etc. \cite{Ultra1,Ultra2,haug,kuhn}. 
However, in undoped quantum wells the lowest electronic excitations are high 
energy interband transitions that react almost instantaneously to the 
photoexcited carriers \cite{louie}. The ground state can then be considered 
as rigid, providing only the band structure and dielectric screening. 
Consequently, Coulomb correlations only occur among 
{\em photoexcited} carriers. In contrast, in doped quantum wells, the presence of a 
2DEG leads to strong Coulomb correlations in the ground state itself, 
resulting in long-range charge and spin order at sufficiently low temperatures.
This order determines the 2DEG reaction to photoexcitation.
Therefore, one must distinguish the effects of this cold 2DEG
from those due to the nonequilibrium photoexcited electron gas. 
The differences manifest themselves in the ultrafast nonlinear response.

\begin{figure*}
\centerline{
\hbox{\psfig{figure=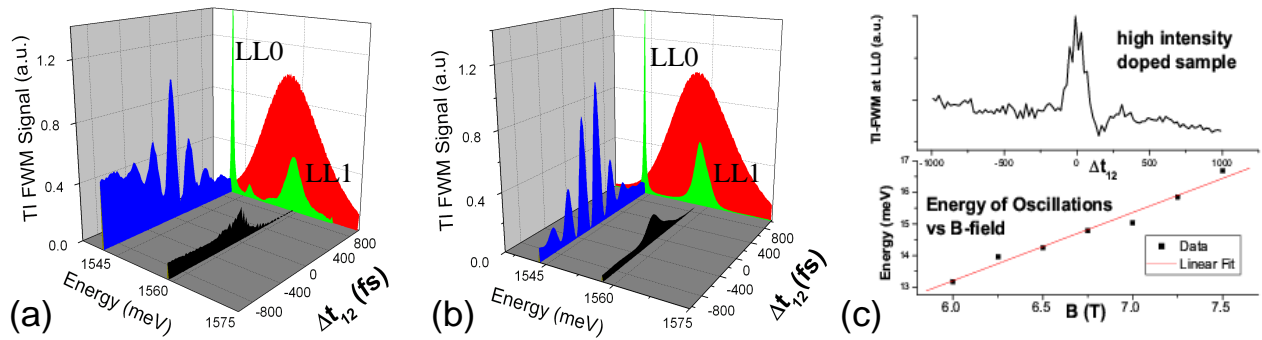,width=17cm}}
}
\caption{ 
Three-pulse FWM signal for the doped quantum well 
along the $\Delta t_{12}$ axis 
($\Delta t_{13}=0$) for low excitation intensity
(a) experiment and (b) theory.
(Backpanel: Linear absorption and optical pulse.)
(c) Top panel: high excitation intensity of doped quantum well along the
$\Delta t_{12}$ axis.
Bottom panel: B-field dependence of the oscillation frequency.}   
\end{figure*}

In this letter we use three-pulse FWM spectroscopy to probe
simultaneously the intra-- and inter--band coherent 
dynamics of the quantum Hall 
system. For low temperatures, weak photoexcitation, and high magnetic 
fields, we observe striking qualitative differences in the coherent 
response of doped and undoped quantum wells. We see a large off-resonant 
signal from the lowest Landau level (LL0), with strong oscillations 
versus pulse time delay in Fig.1a. 
 Using a microscopic 
theory \cite{kara,pera-05}, 
we show that this LL0 signal  is dominated by the interference between 
many--particle 
coherences created by the 
interaction of photoexcited 
carriers with the 2DEG. Photoexcited excitons (X) scatter to X+MP states (i.e. 
states with a photoexcited exciton and a MP excitation of the 2DEG). Within 
ultra-short time scales, this scattering creates a coherence between the
above states - a X$\leftrightarrow$X+MP many--particle coherence. We show that the 
oscillations are 
due to the interference between FWM contributions from these coherences and put
an upper bound on their dephasing rate. The presence of 
the X$\leftrightarrow$X+MP coherences indicates a breakdown of the 
semiclassical 
(Boltzmann) picture \cite{Ultra1,Ultra2,haug} of irreversible scattering.

In our experiment, we excite the quantum Hall system with three 100 fs 
$\sigma_+$-polarized 
pulses along directions ${\bf k}_1$, ${\bf k}_2$, and ${\bf k}_3$.
Pulses ${\bf k}_1$ and ${\bf k}_2$ (${\bf k}_3$) are separated 
by a time delay $\Delta t_{12}$ ($\Delta t_{13}$), where pulse 
${\bf k}_1$ arrives first for negative values of the delay. 
The FWM response is obtained in the background-free direction 
${\bf k}_1+{\bf k}_2-{\bf k}_3$. Using an interference filter of bandwidth
2meV,
we spectrally resolve the response so as to separate out the
contribution from each Landau level. We then measure the intensity from 
each Landau level as a function of the time delays. In particular, we 
measure along the $\Delta t_{12}$ axis ($\Delta t_{13}=0$) or the $\Delta t_{13}$ 
axis ($\Delta t_{12}=0$). For the $\Delta t_{13}$ axis, pulses ${\bf k}_1$ and 
${\bf k}_2$ arrive together. In this case, the physics is similar to 2-pulse FWM 
\cite{from-02}, where ${\bf k}_1$ and ${\bf k}_2$ are degenerate. 
The $\Delta t_{13}$ axis thus reflects the dynamics 
of the optical  polarizations (inter-band) \cite{Ultra1}. On the other 
hand, the $\Delta t_{12}$ axis gives access to new dynamics, like the temporal 
evolution of the X$\leftrightarrow$X+MP coherence (intra-band).

We investigated a modulation-doped quantum well structure consisting of 10 
periods of 12 nm GaAs wells and 42 nm AlGaAs barrier layers with 
Si doped at their centers. The doped carrier (2DEG) density was
2.1$\times$10$^{11}$ cm$^{-2}$ and the low temperature mobility of the 
sample was $\sim 10^5$ cm$^2$/Vs. The sample was kept at $1.5-4^\circ$K 
in a split-coil magneto-optical-cryostat. A perpendicular magnetic 
field (B = 0--7 T) was applied along the 
growth direction of the quantum well. The bulk of our measurements were 
performed at B = 7 T (filling factor $\nu=1.3$).
To isolate the effects of the interactions,
we measured the 
FWM signal from the first Landau level (LL0) while largely  
exciting the second Landau level (LL1) (LL1:LL0 population ratio 
at least 10:1, see back panel of Fig. 1a).  
We then compared the LL0 signal from
the doped quantum well to a similar undoped one (without a 2DEG).
We also compared low and high intensity measurements.  
For low intensity, the photoexcited carrier density (5$\times$10$^9$ cm$^{-2}$) 
was kept much smaller than the 2DEG density
in order to weakly perturb the quantum Hall system. 
For high intensity, the two densities were comparable,   
so the photoexcited carrier contribution 
was strong. 

For large LL1/LL0 photoexcitation ratio, the phase space filling 
contribution to LL0 is very small. The LL0 signal then comes from  LL1-LL0
coupling due to (i) exciton-exciton interactions \cite{undoped}, 
(ii) inter-Landau level coherences of the {\em photoexcited} carriers,
and (iii) X-2DEG interactions. Only the first two contribute in  
undoped semiconductors. Since the LL0 signal in the undoped quantum well is 
small (Fig. 2b), these contributions are weak. In contrast, the doped 
quantum well shows a large LL0 signal (Figs. 1a, 2a). This signal diminishes 
for small magnetic fields, $\nu>2$.
It is also small for high intensities, when the photoexcited carriers dominate
over the 2DEG (Fig. 2c).   
 We therefore conclude that our LL0
signal is due to LL0-LL1 coupling via  X-2DEG interactions.

\begin{figure*}
\centerline{
\hbox{\psfig{figure=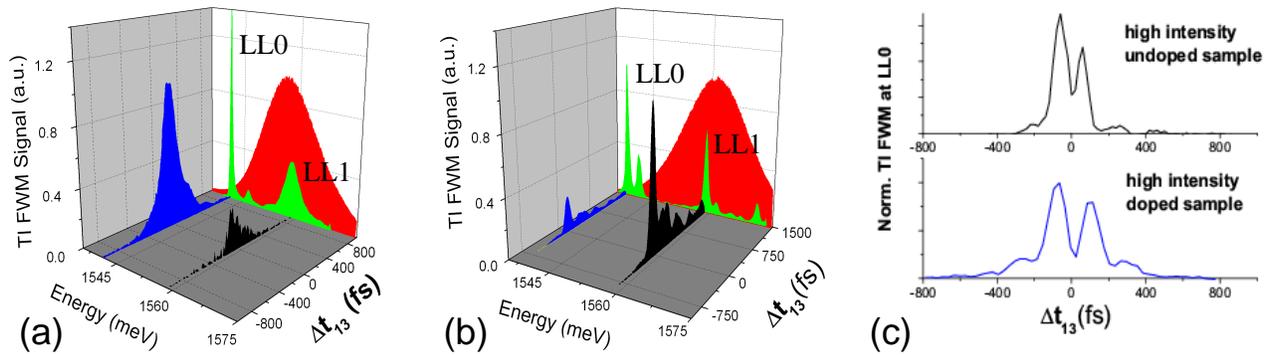,width=17cm}}
}
\caption{ 
Three-pulse FWM signal for low excitation intensity along
the $\Delta t_{13}$ axis ($\Delta t_{12}=0$)
(a) doped and (b) undoped quantum well.
(Backpanel: Linear absorption and optical pulse.)
(c) Comparison of doped and undoped quantum wells for high excitation intensity 
along the $\Delta t_{13}$ axis.}   
\end{figure*}

Along the 
$\Delta t_{13}$ axis, 
we observe striking differences 
 between the doped and undoped quantum well
similar to two-pulse FWM \cite{from-02}. 
However, three-pulse FWM along the $\Delta t_{12}$ axis 
provides new information about the dynamics of X$\leftrightarrow$X+MP coherences
that is not accessible with two-pulse FWM. Fig. 1a shows the FWM signal in the 
doped quantum well along the $\Delta t_{12}$ axis for low photoexcitation, 
high magnetic field, and low temperature. Besides the large transfer of signal 
strength to LL0, we observe strong $\Delta t_{12}$ oscillations only at LL0. 
There are no oscillations at LL1 or in the $\Delta t_{13}$ axis. The oscillation 
frequency is comparable to the inter-Landau level energy spacing and increases 
linearly with the B-field (bottom panel of Fig. 1c). The oscillation decay rate 
for both positive and negative $\Delta t_{12}$ is comparable to the sum 
of the LL0 and LL1 dephasing rates extracted from Fig. 2a. With 
increasing photoexcitation intensity, these $\Delta t_{12}$ 
oscillations disappear quickly, even before the decay of the overall 
signal changes significantly (top panel of Fig. 1c).

{\em Theory:} To understand the non-linear response of the quantum Hall system, we use a 
model derived from the many-body theory of Ref. \cite{pera-05}. This theory 
goes beyond the Dynamics Controlled Truncation Scheme (DCTS) \cite{axt-96}
used to study correlations in undoped semiconductors. 
The DCTS relies on the correspondence between electron-hole pairs 
created or destroyed and the sequence of photons absorbed or emitted.
However, if carriers are present in the system before excitation (e.g. 
doped quantum well), this correspondence breaks down and the DCTS fails. 
We extend the DCTS by tracking only the photoexcited holes created
(destroyed) with the photons absorbed (emitted). Thus
we account for the presence of the 2DEG prior to photoexcitation and 
can treat the X-2DEG interactions.
There is a further challenge in studying the optical response of a 2DEG.
In an undoped system, the excited states can be written as products of 
distinct phonon and X states. However, in the doped system,
both the 2DEG excitations and the Xs are made up of electrons. Thus exchange 
effects complicate a simple factorization. We overcome this difficulty
by introducing a basis of correlated X-2DEG states \cite{pera-05}. 

The third order optical polarization is determined by phase space filling 
and an interaction induced contribution. By minimizing the optical pulse 
overlap at LL0, phase space filling contributions at LL0 are suppressed. 
The interaction induced contributions are described by a density matrix 
$\langle \hat{Y} \rangle$ \cite{kara,pera-05}. 
This includes exciton-exciton interactions, 
which occur in the undoped quantum wells as well, and X-2DEG interactions. 
In Refs. \cite{kara,pera-05} we decomposed $\langle \hat{Y} \rangle$ 
into three parts: {\bf (i)} coherent exciton-exciton interactions as in 
the undoped quantum well; {\bf (ii)} contributions from intraband coherences 
and densities, including the X$\leftrightarrow$X+MP coherences; {\bf (iii)} 
a correlated contribution governed by the time evolution of the X+MP states. 
The latter dephases rapidly and mainly contributes to the LL1 exciton 
linewidth \cite{from-02,kara}.

For simplicity we consider a spin-$\uparrow$ polarized ground state 2DEG 
(realized for $\nu$=1). We then excite spin-$\downarrow$ electrons with 
$\sigma_+$ light. Due to quantitative uncertainties arising from disorder, 
valence band mixing, and higher Landau levels, we determined the independent 
parameters by fitting to the experimental linear absorption (back panel of 
Fig. 1a). Our conclusions are not sensitive to their precise values.
 
The X$\leftrightarrow$X+MP coherence that dominates our results, denoted by $M_0$
from now on, 
is created by the scattering of the LL1 exciton to LL0 with the emission of 
a MP. Note that, due to the degeneracy of the LL1 and LL0+MP states, 
the coherence $M_0$ has small energy compared to the inter-Landau level spacing. 
It can be created by pulses 1-3 ($M_0^{13}$) or pulses 2-3 ($M_0^{23}$). 
The corresponding resonant FWM contribution to LL0 in our calculation comes 
from $P_0M_0^*$, where $P_0$ is the LL0 polarization created by the third pulse 
to probe the $M_0$ coherence \cite{pera-05}. Details of the calculation will be 
presented elsewhere.

{\em Origin of oscillations:} 
Fig. 3a shows a schematic describing the contribution from the $M_0^{13}$ 
coherence. Pulses ${\bf k}_1$ and ${\bf k}_3$ arrive simultaneously 
in the sample to create a density of excitons in LL1. These excitons scatter 
into LL0 with the excitation of a MP, thereby creating the coherence $M_0^{13}$. 
This coherence evolves for a time $|\Delta t_{12}|$, accumulating negligible 
phase due to the small $M_0$ energy. It is then probed by a $P_0$ polarization 
created by ${\bf k}_2$, resulting in a FWM signal in ${\bf k}_1+{\bf k}_2-{\bf k}_3$. 
Due to the symmetry of ${\bf k}_1$ and ${\bf k}_2$ in the 
${\bf k}_1+{\bf k}_2-{\bf k}_3$ signal, we also have a process where ${\bf k}_2$ 
and ${\bf k}_3$ create the coherence $M_0$ (i.e. $M_0^{23}$). This is then probed 
by ${\bf k}_1$. However, one must keep track of the time delays. Thus, as shown in 
Fig. 3b, now ${\bf k}_1$ and ${\bf k}_3$ arrive together with ${\bf k}_3$ 
contributing a LL1 polarization and ${\bf k}_1$ contributing a LL0 polarization. 
These polarizations evolve in the sample for a time $|\Delta t_{12}|$ and 
accumulate a phase of $(\omega_0-\omega_1)\Delta t_{12}$ (where $\hbar\omega_n$ is 
the energy of the $n^{th}$ Landau level). Pulse ${\bf k}_2$ then creates the $M_0^{23}$ 
coherence with the decaying LL1 polarization from ${\bf k}_3$. $M_0^{23}$ is 
instantaneously probed by the decaying LL0 polarization created earlier by ${\bf k}_1$, 
resulting in a FWM signal with the accumulated phase $(\omega_0-\omega_1)\Delta t_{12}$. 
The first process
(due to $M_0^{13}$)
 contributes to the FWM signal 
for $|\Delta t_{12}|$ shorter than 
the decay time  of $M_0^{13}$, while the second process 
(due to $M_0^{23}$) contributes  for $|\Delta t_{12}|$ 
shorter  than the polarization decay times. 
Within the shortest time interval of the two, 
 the contributions from $M_0^{13}$ and $M_0^{23}$ 
will interfere with each other, resulting in oscillations at the inter-Landau level 
frequency $\omega_0-\omega_1$ along the $\Delta t_{12}$ axis.
These do not contribute along the $\Delta t_{13}$ axis,
or in two--pulse FWM \cite{from-02}.  
 Fig. 4 shows the 
numerical calculation of the $M_0^{13}$, $M_0^{23}$ contributions and the interference 
between the two when considered together ($M_0$). The signal from $M_0$ dominates 
the full FWM signal. Contributions from phase space filling and X-X interactions are 
negligible. Fig. 1b shows the full numerical calculation which reproduces well the 
experimental features in the coherent regime. 

\begin{figure}
\centerline{
\hbox{\psfig{figure=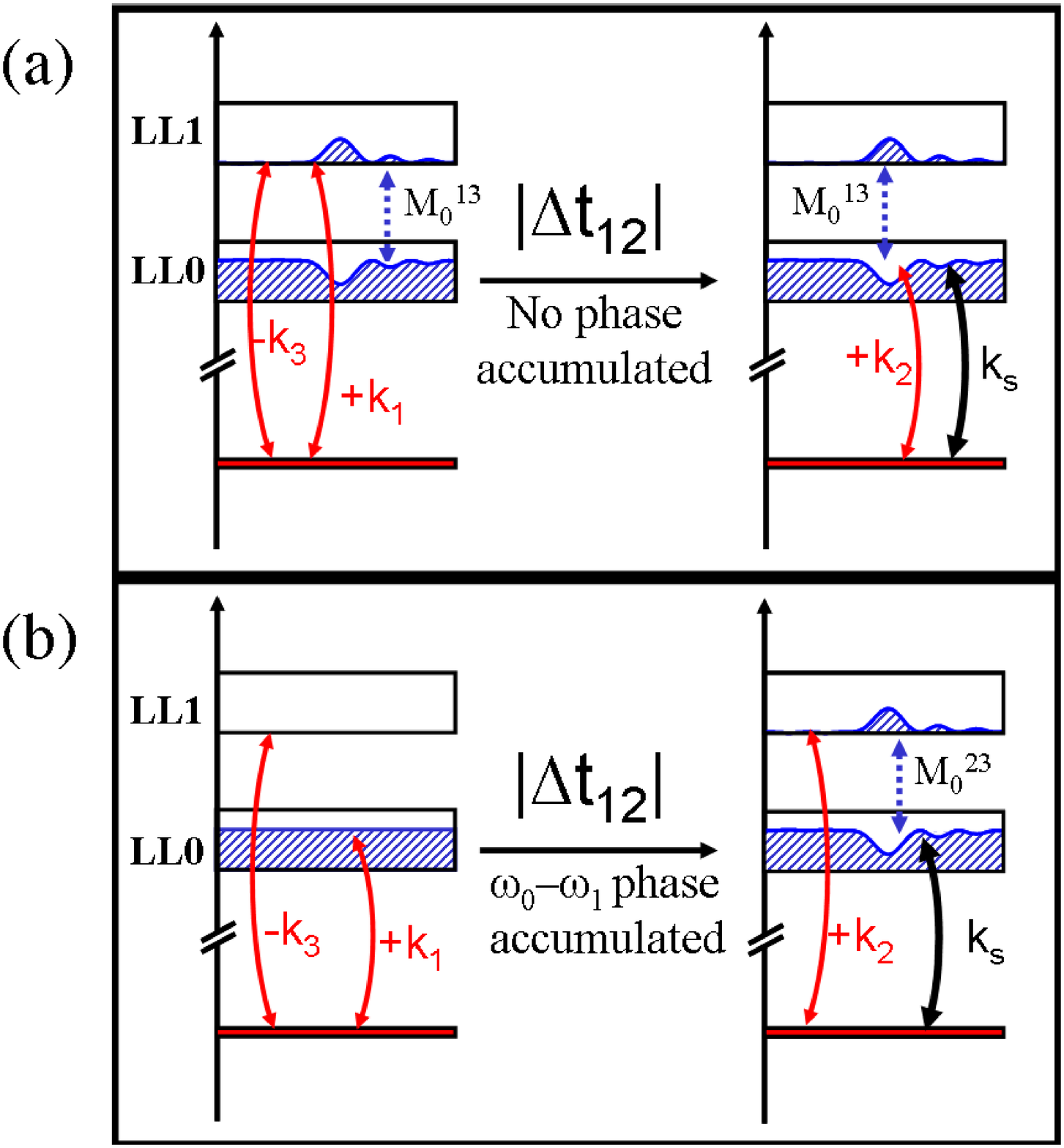,width=6cm}}
}\caption{ 
Third-order process contributing to the FWM signal 
(in the direction ${\bf k}_s={\bf k}_1+{\bf k}_2-{\bf k}_3$) due to
(a) $M_0^{13}$
(b) $M_0^{23}$.
 } 
\end{figure} 
\begin{figure}
\centerline{
\hbox{\psfig{figure=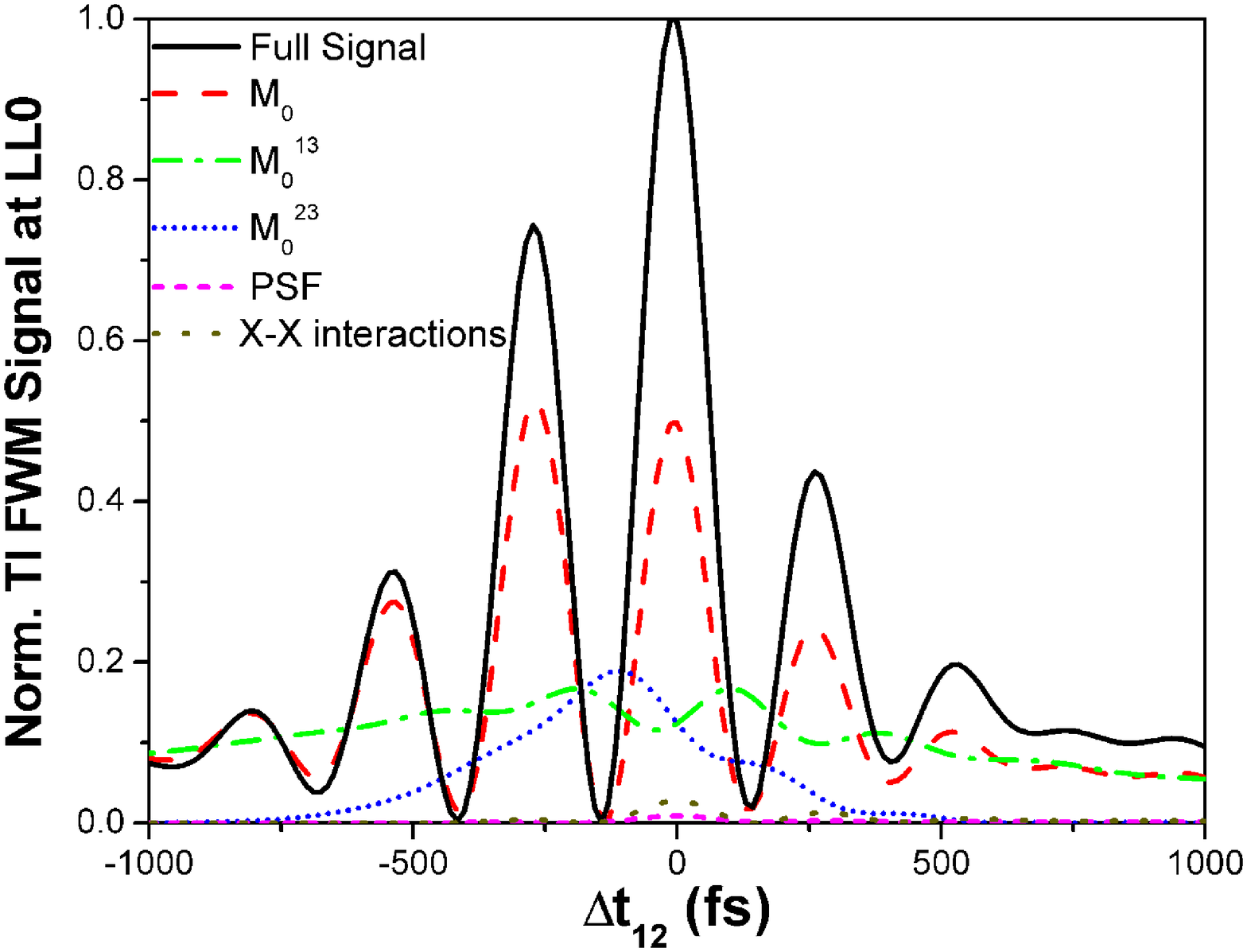,width=7cm}}
}\caption{ 
Numerical calculation of the LL0 FWM signal for the $\Delta t_{12}$ axis
due to different processes.
}
\end{figure}

To interpret the decay of the above oscillations, we assign effective dephasing
rates $\Gamma_0$  and $\Gamma_1$ to the LL0 and LL1 polarizations and assume a 
$M_0$ dephasing rate of $\gamma_0$. The decay of oscillations for $\Delta t_{12}>0$ 
is $\Gamma_0+\Gamma_1$, whereas for $\Delta t_{12}<0$ we get a decay of
$\Gamma_0+\Gamma_1+\gamma_0$. As the experiment shows, the decay of oscillations 
is nearly equal for positive and negative $\Delta t_{12}$. Thus, 
$\gamma_0\ll \Gamma_0+\Gamma_1$, i.e. the X$\leftrightarrow$X+MP coherence
dephases over a time interval of $1/\gamma_0\gg 1/(\Gamma_0+\Gamma_1)\sim 300$ fs.

In conclusion, by using three-pulse FWM spectroscopy, we access for the first tme 
the full coherent 
dynamics of the quantum Hall system. We see a large off-resonant signal with 
strong oscillations only along the $\Delta t_{12}$ axis. Using a microscopic many-body 
theory, we show that the signal is due to many-particle coherences created via the 
non-instantaneous interactions of photoexcited carriers and MPs. The oscillations 
are due to the interference of different FWM contributions of these coherences.
Finally, we put an upper bound to the decay rate of the X$\leftrightarrow$X+MP 
coherence. The combination of ultrafast non-linear spectroscopy and quantum Hall 
physics initiates a new field of coherent quantum Hall dynamics.

We thank N.A. Fromer, S. Cundiff and J. Wang for discussions. 
This work was supported by the Office of Basic Energy Sciences of the
US Department of Energy, the France-Berkeley Foundation, and by the EU
Research Training Network HYTEC.

\end{document}